\documentclass{appolb}
\usepackage{epsfig}

\begin{document}

\title{Clusters in light stable and exotic nuclei}
%\thanks{Invited talk presentted at the EXON2018 International Conference}

\author{C. Beck$^a$
\address{
$^a$D\'epartement de Recherches Subatomiques, Institut Pluridisciplinaire 
Hubert Curien, IN$_{2}$P$_{3}$-CNRS and Universit\'e de Strasbourg - 23, rue 
du Loess BP 28, F-67037 Strasbourg Cedex 2, France\\
E-mail: christian.beck@iphc.cnrs.fr\\}
}

\maketitle

\begin{abstract}

Since the discovery of molecular resonances in $^{12}$C+$^{12}$C 
in the early sixties a great deal of research work has been undertaken to study 
$\alpha$-clustering. Our knowledge on physics of nuclear molecules has 
increased considerably and nuclear clustering remains one of the most fruitful 
domains of nuclear physics, facing some of the greatest challenges and 
opportunities in the years ahead. Occurrence of ``exotic" shapes and 
Bose-Einstein Condensates in light $\alpha$-cluster nuclei are investigated. 
Various approaches of superdeformed/hyperdeformed shapes associated with 
quasimolecular resonant structures are discussed. The astrophysical reaction 
rate of $^{12}$C+$^{12}$C is extracted from recent fusion measurements at deep 
subbarrier energies near the Gamov window. Evolution of clustering from 
stability to the drip-lines is examined.

\end{abstract}

%\bodymatter

\newpage

\section{Introduction}
\label{sec:1}

In the last decades, one of the greatest challenges in nuclear science is the
understanding of the clustered structure of nuclei from both experimental 
and theoretical perspectives
\cite{Cluster1,Cluster2,Cluster3,Correlations,Papka,Oertzen06,Horiuchi,Freer18}. 
Our knowledge on physics of nuclear molecules has increased considerably and 
nuclear clustering remains one of the most fruitful domains of nuclear physics.
Fig.~1 summarizes the different types of clustering~\cite{Correlations}: most 
of these structures were investigated in an experimental context by using 
either some new approaches~\cite{Papka} or developments of older methods. 
The search for resonant structures in the excitation functions for various 
combinations of light $\alpha$-cluster ($N$=$Z$) nuclei in the energy regime 
from the Coulomb barrier up to regions with excitation energies of 
$E_{x}$=20$-$50~MeV remains a subject of contemporary debate. 

\begin{figure}[th]
\centerline{\psfig{figure=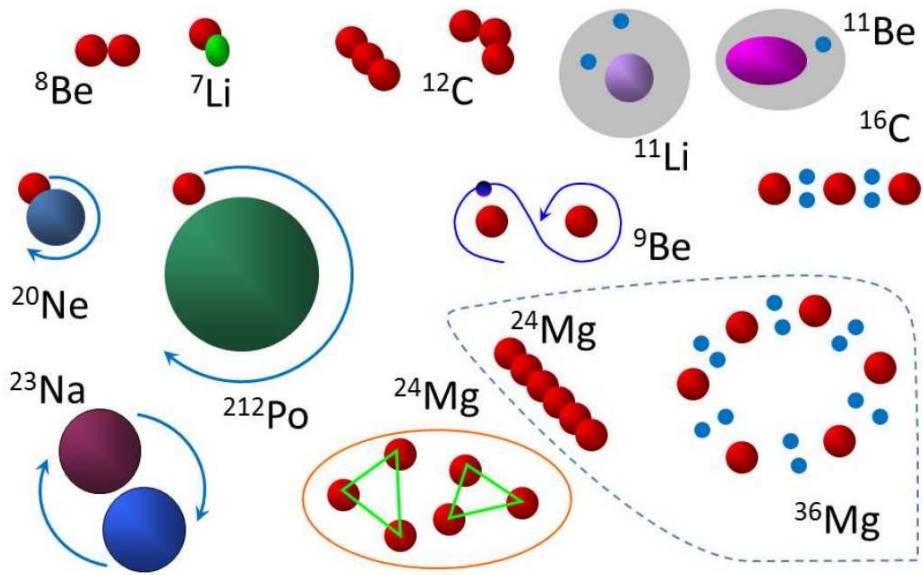,width=12.5cm,height=8.0cm}}
%\vspace*{8pt}
\caption{\label{fig1} 
Different types of clustering behaviour identified in nuclei~\cite{Correlations},
from small clusters outside a closed shell, to complete condensation into 
$\alpha$ particles, to halo nucleons outside of a normal core, have been
discussed the last two or three decades. This figure was adapted from Fig.~1
of Ref. \cite{Correlations} courtesy from Wilton Catford}
\vspace*{-10pt}
\end{figure}

\smallskip

The question of how nuclear molecules may reflect continuous transitions from 
scattering states in the ion-ion potential to true cluster states in the 
compound systems is still unresolved. Clustering in light $\alpha$-like nuclei is 
observed as a general phenomenon at high excitation energy close to the 
$\alpha$-decay thresholds \cite{Correlations}. 
This exotic behavior has been perfectly illustrated 50 years ago by the famous 
''Ikeda-diagram" for $N$=$Z$ nuclei~\cite{Ikeda}, which has been 
modified and recently extended by von Oertzen 
\cite{Oertzen06} for neutron-rich nuclei, as shown in the
left panel of Fig.~2. Despite the early inception of cluster studies, it is only
recently that radioactive ion beams experiments, with great helps from advanced
theoretical works, enabled new generation of studies, in which data with
variable excess neutron numbers or decay thresholds are compared to predictions
with least or no assumptions of cluster cores. Some of the predicted but elusive
phenomena, such as molecular orbitals or linear chain structures, are now
gradually coming to light.  

\smallskip

\begin{figure}[th]
\centerline{\psfig{figure=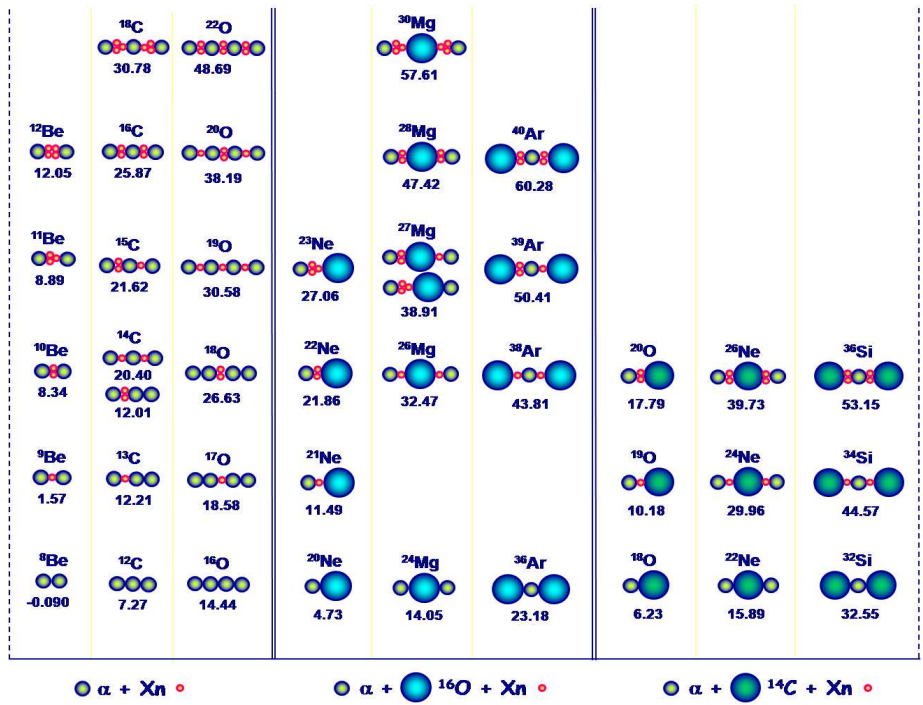,width=12.0cm,height=9.3cm}}
\vspace*{8pt}
\caption{\label{fig2} Schematic illustration of the structures of molecular
shape isomers in light neutron-rich isotopes of nuclei consisting
of $\alpha$-particles, $^{16}$O- and $^{14}$C-clusters plus some
covalently bound neutrons (Xn means X neutrons) \cite{Oertzen06}. The so called 
''Extended Ikeda-Diagram" with $\alpha$-particles (left panel) and 
$^{16}$O-cores (middle panel) can be generalized to $^{14}$C-cluster cores 
(right panel). The lowest line of each configuration corresponds to parts
of the original ''Ikeda-Diagram" \cite{Ikeda}. Threshold energies, dissociating
the ground state into the respective cluter configuration, are given in MeV.
This figure has been adapted courtesy from Wolfram von Oertzen.}
\end{figure}

\newpage

\section{$^{12}$C nucleus ''Hoyle" state and BEC in light nuclei}
\label{sec:2}

The ground state of $^8$Be is the most simple and convincing example
of $\alpha$-clustering in light nuclei as suggested by several theoretical
models and appears naturally in {\it ab initio} calculations 
\cite{Correlations,Freer18}. The picture of the $^8$Be nucleus prediced by the 
No Core Shell model~\cite{Freer18} as being a dumbbell-shaped configuration of 
two $\alpha$ particles closely resembles the superdeformed (SD) shapes known 
to arise in heavier nuclei in the actinide mass region. This dumbbell-like 
structure gives rise to a rotational band, from which the moment of inertia is 
found to be commensurate with an axial deformation of 2:1. According to
the schematic picture of the ''Ikeda-Diagram" \cite{Ikeda} the nuclear cluster 
structure of $^{12}$C may also induce axial deformation close to 3:1 of
a hyperdeformed (HD) shape. The large deformations  of light $\alpha$-conjugate 
nuclei with SD, HD and linear-chain configurations are under discussion.\\
 
The renewed interest in $^{12}$C was mainly focused to a better understanding 
of the nature of the so called ''Hoyle" state \cite{Hoyle54,Freer14}, the 
excited 0$^+$ state at 7.654 MeV that can be described in terms of a bosonic 
condensate, a cluster state and/or a $\alpha$-particle gas \cite{Tohsaki01}. 
The resonant ''Hoyle" state \cite{Hoyle54} is regarded as the prototypical 
$\alpha$-cluster state whose existence is of great importance for the 
nucleosynthesis of  $^{12}$C within stars. Further knowledge of the ''Hoyle" 
state \cite{Hoyle54,Freer14} and its rotational excitations would help not
only to understand the debated structure of the $^{12}$C nucleus in the ``Hoyle
state", but also to determine the high-temperature (T $\approx$ 1 GK) reaction
rate of the triple $\alpha$ process more precisely. The structure of this state has 
been thoroughly investigated with theoretically modelled with both {\it ab 
initio} and cluster models \cite{Correlations,Freer18}. Much experimental progress 
has been achieved recently as far as the spectroscopy of $^{12}$C near and above 
the $\alpha$-decay threshold is concerned~\cite{Kirsebom17}. More particularly, 
the the second 2$^{+}_{2}$ ''Hoyle" rotational excitation in  $^{12}$C has 
been observed \cite{Zimmerman13}. Another experiment~\cite{Marin14} populates 
a new state compatible with an equilateral triangle configuration of three $\alpha$ 
particles. Still, the structure of the ''Hoyle" state remained controversial
as experimental results of its direct decay into three $\alpha$ particles are found 
to be in disagreement until two experiments provided the most precise picture of 
how a $^{12}$C excited state decays into three He 
nuclei~\cite{Kirsebom17,Aquila17,Smith17}.\\

In the study of Bose-Einstein Condensation (BEC), the $\alpha$-particle states 
in light $N$=$Z$ nuclei \cite{Tohsaki01}, are of great interest. the search for 
an experimental signature of BEC in $^{16}$O is of highest priority. 
Furthermore, {\it ab initio} calculations \cite{Correlations,Freer18} predict 
that nucleons are arranged in a tetrahedral configuration of $\alpha$ clusters.
A state with the structure of the ''Hoyle" state \cite{Hoyle54} in 
$^{12}$C coupled to one $\alpha$ particle is 
predicted in  $^{16}$O at about 15.1 MeV (the 0$^{+}_{6}$ state), the
energy of which is $\approx$ 700 keV above the 4$\alpha$-particle 
breakup threshold. However, any state in
$^{16}$O equivalent 
to the ''Hoyle" state \cite{Hoyle54} in $^{12}$C is most certainly 
going to decay exclusively by particle emission with very small 
$\gamma$-decay branches, thus, very efficient
particle-$\gamma$ coincidence techniques~\cite{Papka} will 
have to be used in the near future to search for them. 

\newpage

\section{Nuclear molecules, $^{12}$C+$^{12}$C reaction rate and carbon burning 
in massive stars}
\label{sec:3}

The real link between superdeformation/hyperdeformation (SD/HD), nuclear molecules 
and $\alpha$-clustering \cite{Correlations} is of particular interest, 
since nuclear shapes with major-to-minor axis ratios of 2:1--3:1 have the typical 
ellipsoidal elongation for light nuclei. A further area where electromagnetic 
transitions would be of great interest in support of cluster models is in the 
case of the quasi-molecular resonances observed in the $^{12}$C+$^{12}$C reaction. 
The widths of these resonances were $\approx$ 100 keV, indicating the formation of a 
$^{24}$Mg intermediate system with a lifetime significantly longer than the nuclear 
crossing time. These resonances were subsequently interpreted as $^{12}$C+$^{12}$C 
cluster states. There has been only one valient attempt to directly observe 
transitions in this reaction~\cite{Correlations} focussing on 
transitions between 10$^{+}$ and 8$^{+}$ resonant states at a bombarding energy 
E($^{12}$C) = 32 MeV chosen to populate a known and isolated 10$^+$ resonance. 
However, the measurement reported only an upper limit (for the radiative partial 
width of 1.2 $\pm$ 10$^{-5}$) given the extreme challenges of eliminating all 
background. \\

\begin{figure}[th]
\centerline{\psfig{figure=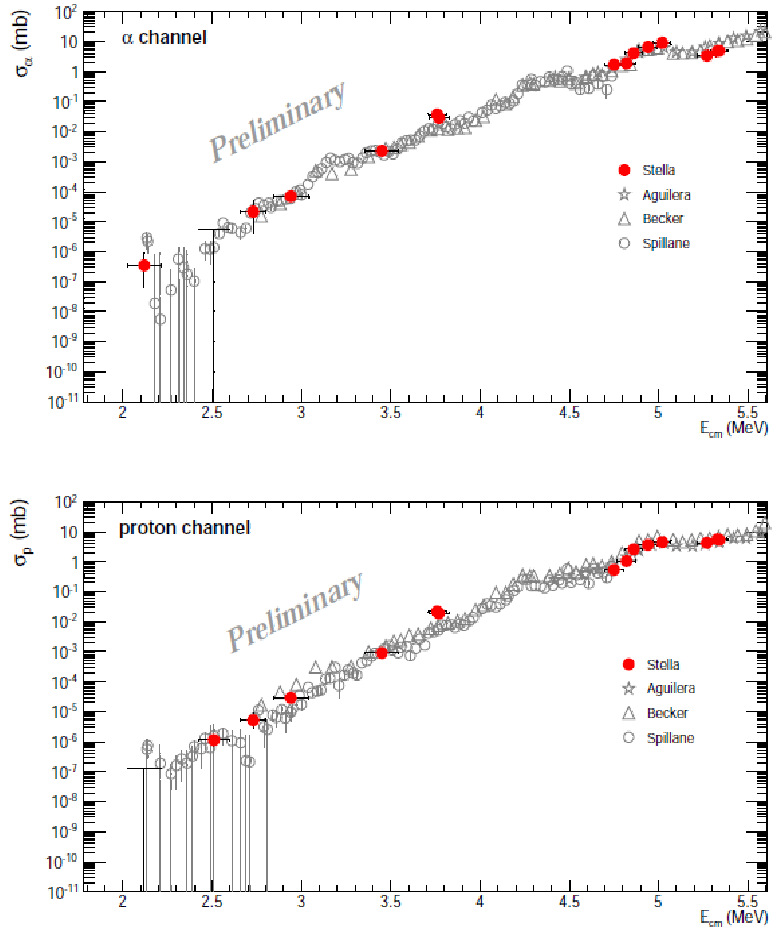,width=10.4cm,height=12.5cm}}
\vspace*{8pt}
\caption{\label{fig3}$^{12}$C+$^{12}$C fusion-evaporation excitation functions
\cite{Fruet2018} ($\alpha$ upper panel and p lower panel) as obtained by
the Stella collaboration~\cite{Heine2018}. Comparisons either with previously data
obtained by H.W. Becker et al.~\cite{Becker} (open triangles), E.F. Aguilera et al. 
\cite{Aguilera} (open stars), and T. Spillane et al.~\cite{Spillane07} (open circles),
respectively, or with the Fowler model~\cite{Fowler} (black dotted line) and
the hindrance model~\cite{Jiang} (red dashed line). Courtesy from G. 
Fruet.}
\end{figure}

\begin{figure}[th]
\centerline{\psfig{figure=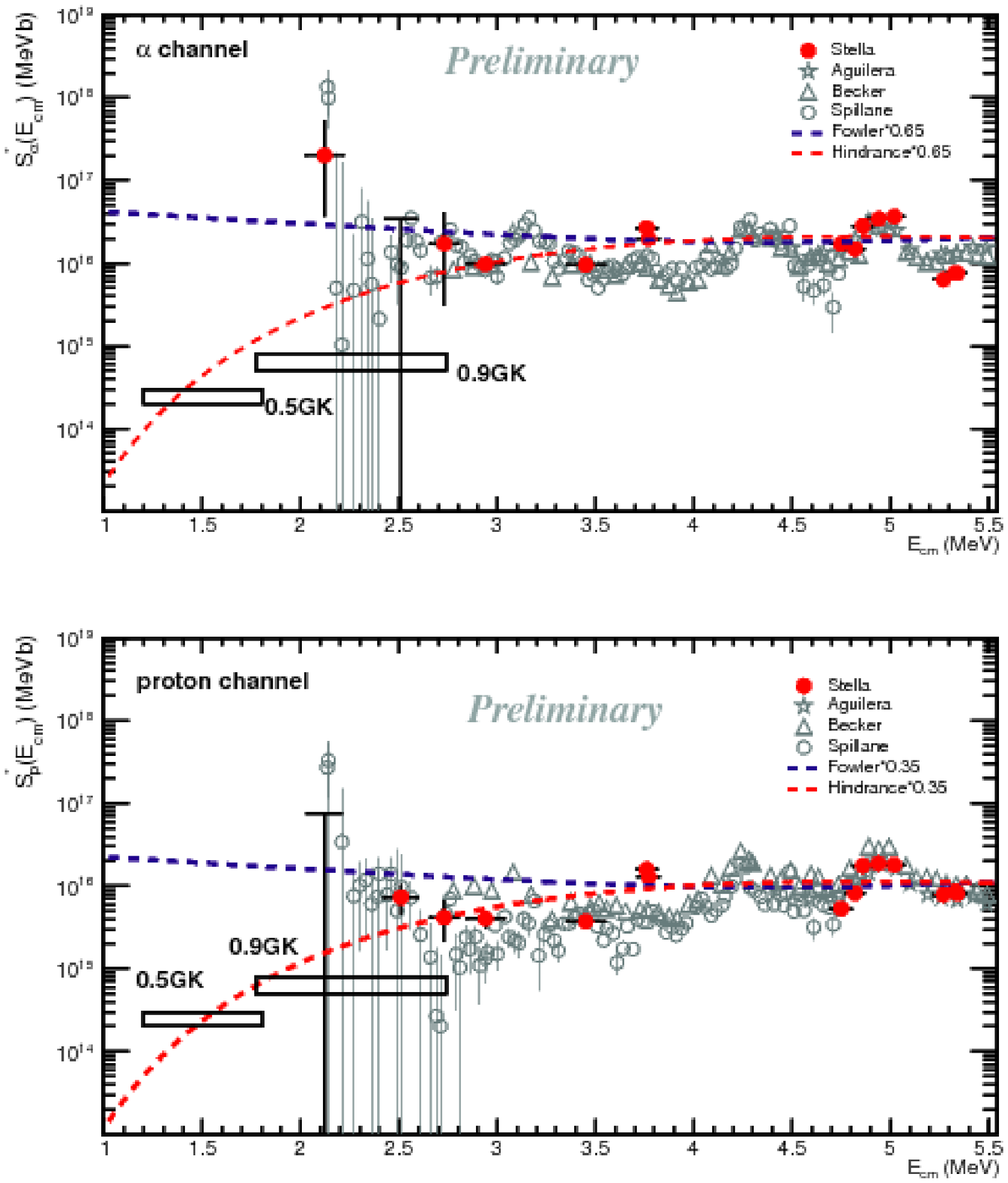,width=10.4cm,height=12.5cm}}
\vspace*{8pt}
\caption{\label{fig3}Modified $^{12}$C+$^{12}$C astrophysical S-factors
\cite{Fruet2018} ($\alpha$ upper panel and p lower panel) as obtained by
the Stella collaboration~\cite{Heine2018}. Comparisons either with previously data
obtained by H.W. Becker et al.~\cite{Becker} (open triangles), E.L. Aguilera et al. 
\cite{Aguilera} (open stars), and T. Spillane et al.~\cite{Spillane07} (open circles),
respectively, or with the Fowler model~\cite{Fowler} (black dotted line) and
the hindrance model~\cite{Jiang} (red dashed line). The two black rectangles indicate
the positions of the Gamow window stellar temperature of T~=~0.9 GK (E$_0$~=~ 2.25
$\pm$ 0.46 MeV) and T~=~0.5 GK (E$_0$~=~1.5 $\pm$ 0.3 MeV). Error barrs correspond to
statistical uncertainties. Courtesy from G. Fruet.}
\end{figure}

The role of cluster configurations in stellar helium burning is well established
and, discussion about the nature and the role of resonance structures that
characterize the low-energy cross section of the $^{12}$C+$^{12}$C
fusion process is underway in recent experimental investigations
\cite{Spillane07,Jiang13,Bucher15,Jiang18,Tumino18a,Zickefoose18,Heine2018,Fruet2018}.
The resonant structures at very low energies have still been identified
as molecular $^{12}$C+$^{12}$C configurations in the $^{24}$Mg compound
nucleus \cite{Correlations,Spillane07}. However, the reaction rate is calculated 
on the basis of an average cross section integrating over the molecular resonance 
components. Indications of possible existence of a pronounced low-energy resonance 
at E$_{cm}$ = 2.14 MeV \cite{Spillane07,Fruet2018} that can only be 
explained by strong $^{12}$C cluster configurations of the corresponding state
in $^{24}$Mg. \\

There have been also predictions based on phenomenological considerations of explosive 
stellar events, such as X-ray superbursts, type Ia supernovae, stellar evolution 
etc..., that suggest a strong $^{12}$C+$^{12}$C cluster resonance around E$_{cm}$ = 
1.5 MeV in $^{24}$Mg that would drastically enhance the energy production and may 
provide a direct nuclear driver for the superburst phenomenon \cite{Wiescher,Kajino}. 
However, no indication for such a state was reported. Much of the data collected to date 
\cite{Spillane07,Becker,Aguilera} are shown in Fig.~3 taken from Ref.~\cite{Fruet2018}. 
First direct $^{12}$C+$^{12}$C measurement~\cite{Spillane07} seemed to indicate such a 
resonance. Recent measurements were performed at deep subbarrier energies using the
newly developed Stella apparatus \cite{Heine2018} associated with the UK FATIMA 
detectors \cite{Fatima} for the exploration of fusion cross sections of astrophysical 
interest \cite{Fruet2018}. Gamma-rays have been
detected in an array of LaBr$_3$ scintillators whereas proton and
$\alpha$-particles were identified in double-sided silicon-strip detectors. A
novel rotating target system has been developed in order to be capable to
sustain high-intensity carbon beams delivered by the Andom\`ede facility of
the University Paris-Saclay and IPN Orsay, France \cite{Andromede}. The particle-$\gamma$
coincidence technique as well as nanosecond timing conditions have been used in
the data analysis in order to minimize the background as much as possible.
Our preliminary results~\cite{Fruet2018} obtained with Stella~\cite{Heine2018} 
confirm the possible occurence of such a resonant structure in the $\alpha$ channel but 
not in the proton channel. \\

At higher energies the $^{12}$C+$^{12}$C cross sections expressed for 
Stella~\cite{Heine2018} in terms of the modified astrophysical S-factor are typically
in fair agreement with those measured at Argonne \cite{Jiang18} with similar coincidence 
techniques \cite{Jiang12}. The comparisons with previous data obtained by Becker et 
al.~\cite{Becker} (open triangles), E. Aguilera et al.~\cite{Aguilera} (open stars), 
and Spillane et al.~\cite{Spillane07} (open circles), respectively, show a perfect 
agreement each other. All sets of chosen data lie in between the Fowler model~\cite{Fowler} 
(black dotted line) and the hindrance model~\cite{Jiang} (red dashed line).\\

Our first conclusions might be summarized such as we confirm
the possible fusion hindrance plus persisting resonances near the Gamow energy window.
Furthermore, the preliminary Stella S-factors appear to be in qualitative agreement
with either classical coupled-channel calculations of Esbensen \cite{Esbensen,Notani} or
more recent theoretical investigations~\cite{Rowley,Alexis,Khoa}. It is not clear that 
recent studies \cite{Tumino18a} using the Trojan Horse Method technique (THM) 
confirmed the cluster level at E$_{cm}$ = 2.1 MeV but rather suggested the existence 
of the predicted state at E$_{cm}$ =  1.5 MeV for the fusion reaction. Such a 
discovery of a low energy $^{12}$C+$^{12}$C cluster state would indeed have 
significant impact on the reaction rate; but some doubts \cite{Mukha18a,Mukha18b} 
have been raised to our attention as far as the validity of the indirect THM is 
concerned \cite{Tumino18a,Tumino18b}. Obviously an experimental confirmation through 
direct fusion studies would be of utmost importance and several experiments are underway
\cite{Fang2017}. On the other hand, it is expected 
that if a strong resonance indeed exists around the Gamow energy, then the theoretical 
structure studies should be able to predict a 0$^+$ excited state of $^{24}$Mg or 
an L~=~0 $^{12}$C+$^{12}$C resonance at this particular energy, which plays about 
the same igniting role as that of the ''Hoyle" state \cite{Hoyle54} in the 
triple-$\alpha$ process of $^{12}$C formation \cite{Freer14}. A multichannel folding
model~\cite{Assuncao13} demonstrates the importance of inelastic channels involving
the ``Hoyle state" especially in the low-energy range relevant in astrophysics in
the vicinity of the Gamow region.\\

\newpage

\section{Clustering in light neutron-rich nuclei}
\label{sec:4}

Clustering is a general phenomenon observed also in nuclei with extra neutrons as 
it is presented in the ''Extended Ikeda-diagram" \cite{Ikeda} proposed by von 
Oertzen \cite{Oertzen06} (see the left panel of Fig.~2). With additional neutrons, 
specific molecular structures appear with binding effects based on covalent 
molecular neutron orbitals. In these diagrams $\alpha$-clusters and 
$^{16}$O-clusters (as shown by the middle panel of the diagram of Fig.~2) are the 
main ingredients. Actually, the $^{14}$C nucleus may play similar role in 
clusterization as the $^{16}$O one since it has similar  properties as a cluster: 
i) it has closed neutron p-shells, ii) first excited states are well above 
E$^{*}$ = 6 MeV, and iii) it has high binding energies for $\alpha$ particles.\\

The possibility of extending molecular structures from dimers (Be isotopes) to 
trimers \cite{Oertzen06} has been investigated in detail for C and O isotopes
\cite{Oertzen14}. For C isotopes the neutrons would be exchanged between the three centers
($\alpha$ particles). It is possible that the three $\alpha$-particle configuration 
can align themselves in a linaer fashion, or alternative collapse into a triangle 
arrangment - in either case the neutrons being localised across the three centers. 
Possibly the best case for the linear arrangement is $^{16}$C. \\

A general picture of clustering and molecular configurations in light nuclei can also 
be drawn from the detailed investigation of the light O isotopes \cite{Oertzen14}. The 
bands of $^{20}$O \cite{Oertzen14} compared with the ones of $^{18}$O clearly 
establishes parity inversion doublets predicted by both the Generator-Coordinate-Method
(GCM) and the Antisymmetrized Molecular Dynamics (AMD) \cite{Horiuchi} calculations for 
the $^{14}$C--$^6$He cluster and $^{14}$C--2n--$\alpha$ molecular structures. The 
corresponding moments of inertia are suggesting large deformations for the cluster 
structures. \\

We may conclude that the  reduction of the moments of inertia of the lowest bands of 
$^{20}$O is consistent with the assumption that the strongly bound $^{14}$C nucleus 
having equivalent properties to $^{16}$O, has a similar role as $^{16}$O in relevant, 
less neutron rich nuclei. Therefore, the ''Ikeda-Diagram \cite{Ikeda} and the "extended 
Ikeda-Diagram" consisting of $^{16}$O cluster cores with covalently bound neutrons 
must be further extended to include also the $^{14}$C cluster cores as illustrated 
in Fig.~2. \\

\newpage

\section{Summary and outlook}

The link of $\alpha$-clustering, quasimolecular resonances and extreme deformations 
(SD, HD etc...) has been discussed. Several examples emphasize the general connection 
between molecular structure and deformation effects within {\it ab initio} models and/or 
cluster models~\cite{Freer18}. We have also presented the BEC picture of light (and 
medium-light) $\alpha$-like nuclei that appears to be an alternate way of understanding 
most of properties of nuclear clusters~\cite{Correlations}. New results regarding cluster 
and molecular states in neutron-rich oxygen isotopes in agreement with AMD predictions 
are summarized~\cite{Oertzen14}. Consequently, the ''Extended Ikeda-diagram" has been 
further modified for light neutron-rich nuclei by inclusion of the $^{14}$C cluster, 
similarly to the $^{16}$O one. Marked progress has been made in many traditional and 
novels subjects of nuclear cluster physics and astrophysics (stellar He burning
\cite{Heine2018,Fruet2018,Wiescher,Kajino}). \\

The developments in these subjects 
show the importance of clustering among the basic modes of motion of nuclear many-body 
systems. All these open questions will require precise coincidence measurements 
\cite{Papka} coupled with state-of-the-art theory \cite{Correlations,Horiuchi,Freer18}.\\

\newpage

\section{Dedication and acknowledgements}

This written contribution is dedicated to the memory of my friends Alex Szanto de Toledo,
Valery Zagrebaev, Walter Greiner and Paulo Gomes who unexpectelly passed away since early 
2015. I am very pleased to first acknowledge Walter Greiner for his continuous support of 
the cluster physics \cite{Greiner95,Strasbourg,Zagrebaev10,Poenaru10,Greiner08}. 
I would like to thank Christian Caron (Springer) for initiating in 2008 the series 
of the three volumes of \emph{Lecture Notes in Physics} entitled "Clusters in Nuclei" and 
edited between 2010 and 2014 \cite{Cluster1,Cluster2,Cluster3}. All the 37 authors of the 19 
chapters of these volumes are warmly thanked for their
fruitfull collaboration during the course of the project which is still in
progress \cite{Papka,Horiuchi,Oertzen14,Zagrebaev10,Poenaru10,
Gupta10,Kanada10,Oertzen10a,Ikeda10,
Descouvemont12,Nakamura12,Baye12,Adamian12,Yamada12,Deltuva14,Jenkins14,Zarubin14,Simenel14,Kamanin14}. Thanks 
also to Udo Schroeder for the edition of the volume "Nuclear Particle Correlations and Cluster 
Physics" that inspired \cite{Correlations} so much several aspects presented at the EXON2018 
Symposium in September 2018 as well as at the ISPUN2017 Symposium held in September 2017 in Halong Bay, Vietnam.
Special thanks to all the members of the Stella collaboration \cite{Heine2018}, in
particular Sandrine Courtin, Guillaume Fruet, Marcel Heine, Mohamad Moukaddam, Dominique 
Curien, et al. from the IPHC Strasbourg, Serge Della Negra (Androm\`ede accelerator
\cite{Andromede}) 
{\it et al.} from IPN Orsay, David 
Jenkins, Paddy Regan (UK FATIMA collaboration \cite{Fatima}) {\it et 
al.} from the UK and Christelle Stodel from GANIL. 
The Stella collaboration is supported by the french {\it ``Investissements d'avenir"} program, the
University of Strasbourg {\it ``IdEx Attractivity"} program and the USIAS, Strasbourg, France.  
Finally, Dao Khoa, Le Hoang Chien, Alexis Diaz-Torres, Cheng-Lie Jiang, Akram 
Mukhamedzhanov and Xiadong Tang are acknowledged for their carefull reading of the manuscript.

\newpage

%\section*{References}

%\begin{thebibliography}{99}

\begin{center}

\end{center}
\end{document}